\documentclass[acmsmall,nonacm]{acmart}

\usepackage[ruled,vlined]{algorithm2e}
\usepackage{hyperref}
\usepackage{siunitx}

\AtBeginDocument{%
  }

\begin{document}

\title{FlipHash: A Constant-Time Consistent Range-Hashing Algorithm}

\author{Charles Masson}
\affiliation{%
  \institution{Datadog}
  \city{New York}
  \state{New York}
  \country{USA}
}
\email{charles@datadog.com}
\author{Homin K. Lee}
\affiliation{%
  \institution{Datadog}
  \city{New York}
  \state{New York}
  \country{USA}
}
\email{homin@datadog.com}


\begin{abstract}
  Consistent range-hashing is a technique used in distributed systems, either directly or as a subroutine for consistent hashing, commonly to realize an even and stable data distribution over a variable number of resources. We introduce FlipHash, a consistent range-hashing algorithm with constant time complexity and low memory requirements. Like Jump Consistent Hash, FlipHash is intended for applications where resources can be indexed sequentially. Under this condition, it ensures that keys are hashed evenly across resources and that changing the number of resources only causes keys to be remapped from a removed resource or to an added one, but never shuffled across persisted ones. FlipHash differentiates itself with its low computational cost, achieving constant-time complexity. We show that FlipHash beats Jump Consistent Hash's cost, which is logarithmic in the number of resources, both theoretically and in experiments over practical settings.

\end{abstract}




\maketitle

\section{Introduction}

In distributed databases, data is often required to be horizontally partitioned into a number of resources, e.g., shards, that can be indexed by a range of integers from $0$ to $n-1$, that varies over time to ensure that the overall system scales with the amount of data. As the number of shards changes, it is desirable to keep the amount of data that is shuffled across shards as small as possible. Specifically, it is preferable that as one shard is added, a minimal $1/(n+1)$ share of the data is redistributed from each of the previously existing shards to the newly added shard to maintain the balanced distribution of data across shards. In addition, no data should be unnecessarily shuffled across the previously existing shards. Given that mapping keys to shards is an operation that has to be performed on every insert, we want this mapping to be as computationally cheap as possible. Consistent range-hashing is a technique that provides a solution to this problem, and we introduce a new algorithm that addresses these particular constraints.

\subsection{Background}

\emph{Consistent hashing}~\cite{DBLP:conf/stoc/KargerLLPLL97} aims at mapping keys to a variable set of nodes or resources (e.g., web servers, shards) while ensuring both monotonicity and balance. \emph{Monotonicity}~\cite{DBLP:conf/stoc/KargerLLPLL97} \cite{DBLP:journals/corr/AppletonO15} \cite{DBLP:journals/corr/LampingV14} \cite{DBLP:journals/corr/abs-2107-07930} \cite{DBLP:conf/sebd/ColuzziBL23}, also referred to as minimal disruption~\cite{Rendezvous} \cite{DBLP:conf/nsdi/EisenbudYCSKMCC16} \cite{DBLP:journals/ton/MendelsonVBLKO21} \cite{DBLP:journals/corr/abs-2107-07930}, stability~\cite{DBLP:conf/soda/MirrokniTZ18}, resilience to backend changes~\cite{DBLP:conf/nsdi/EisenbudYCSKMCC16}, remapping property~\cite{DBLP:journals/corr/Sackman15}, or simply consistency~\cite{DBLP:journals/tpds/Nakatani21}, is the property of a hash function that minimizes the number of keys being remapped to a different bucket when the set of buckets varies. \emph{Balance}~\cite{DBLP:conf/stoc/KargerLLPLL97} \cite{DBLP:journals/corr/LampingV14} \cite{DBLP:conf/nsdi/EisenbudYCSKMCC16} \cite{DBLP:journals/ton/MendelsonVBLKO21} \cite{coluzzi2024mementohash} \cite{DBLP:conf/sebd/ColuzziBL23}, also referred to as load balancing~\cite{Rendezvous} \cite{DBLP:journals/corr/AppletonO15} \cite{DBLP:conf/soda/MirrokniTZ18}, uniform balance~\cite{DBLP:journals/corr/abs-2107-07930}, uniform load balancing~\cite{DBLP:journals/tpds/Nakatani21} or uniformity~\cite{DBLP:journals/corr/Sackman15}, is achieved when the keys are hashed evenly across buckets.

Consistent hashing is particularly relevant when both the cost of shuffling keys across resources is high and individual resources cannot be overloaded. It is commonly used in the contexts of distributed caching~\cite{DBLP:conf/stoc/KargerLLPLL97}, load balancing~\cite{DBLP:conf/nsdi/EisenbudYCSKMCC16} \cite{DBLP:conf/nsdi/OlteanuAVR18}, and distributed databases~\cite{DeCandia2007} \cite{DBLP:journals/sigops/LakshmanM10}. As such, it has become the increased focus of research over the past decades, and various solutions have been proposed, each with their own trade-offs.

Our focus is on the class of problems that allow for the sequential indexing of the set of resources, from $0$ to some variable integer $n$. We call this \emph{consistent range-hashing}. In other words, resources may be added at will, and some keys will be remapped to them so as to maintain balance. However, arbitrary resource removals are not allowed, and only the last added resource can be removed. Adding this restriction enables the better fulfillment of the monotonicity and balance properties, with computational and memory cost lower than those of the solutions to the more general consistent hashing problem mentioned above. One concrete use of consistent range-hashing is in database applications, where a keyspace needs to be hashed evenly to a variable number of shards that are backed by multiple replicas. This resource redundancy avoids the need for dealing with arbitrary resource loss in the hashing algorithm itself.

\subsection{Related work}

Rendezvous hashing~\cite{Rendezvous} \cite{DBLP:journals/ton/ThalerR98}, also known as highest random weight (HRW) hashing, was invented in 1996. Even though it was not referred to as a consistent hashing algorithm, it strives to map keys to nodes while achieving the same goals of monotonicity and regularity. Keys and nodes are hashed together as pairs to generate weights, and a key is mapped to the node whose pair with the key realizes the highest weight. Because mapping a key to a node requires calculating the weights of the pairs that the key forms with each of the nodes, the evaluation time is $O(n)$, where $n$ is the number of nodes.

Consistent hashing was introduced as such in 1997 by Karger et al.~\cite{DBLP:conf/stoc/KargerLLPLL97} \cite{DBLP:journals/cn/KargerSBBDIKMY99} in the context of distributed web caching. They describe a solution that associates a fixed number of virtual nodes with each node and hashes them to a ring. Keys are also hashed to the ring and are mapped to the next (or nearest) node on the ring, which is done in $O(1)$ time. This solution achieves monotonicity but does not perfectly balance the keyspace across nodes. One measure of load dispersion across nodes that is often used in the context of consistent hashing is the peak-to-average ratio, that is, the ratio between the maximum and the average number of keys mapped to a single node, because it is an indicator of how over-provisioned the nodepool needs to be to accommodate the load of the most loaded node. To achieve a peak-to-average ratio of $1+\epsilon$, $\Theta(\epsilon^{-2}\log n)$ virtual nodes are required, therefore $O(\epsilon^{-2} n\log n)$ memory. The number of virtual nodes needs to be decided beforehand, as later changing it breaks monotonicity. That implies that the number of nodes needs to stay below an initially set upper bound that depends on the number of virtual nodes in order for the peak-to-average load ratio to stay below a certain threshold.

It was later shown~\cite{DBLP:journals/corr/AppletonO15} that load balance can also be improved with \emph{multiprobing}, where a key is hashed multiple times to the ring and mapped to the node that is the nearest to any of those hashes. This solution achieves $1+\epsilon$ peak-to-average ratio with a $O(1/\epsilon)$ hashing time and $O(n)$ memory space. As opposed to virtual nodes, multiprobing allows indefinitely adding nodes without degrading the peak-to-average load ratio.

Noting that some applications can tolerate a small amount of shuffling in the assignments of keys to nodes, Maglev~\cite{DBLP:conf/nsdi/EisenbudYCSKMCC16} offers a way to achieve better load balance to the detriment of strict monotonicity. The ring is sliced into $\Theta(n/\epsilon)$ equal-sized buckets and nodes take turns picking their next preferred bucket from those still available, the order of preference for each node being generated consistently. In addition to giving up on strict monotonicity, this algorithm suffers from a costly $O((n/\epsilon)\log(n/\epsilon))$ update time following the addition or the removal of a node.

So-called perfect consistent hashing~\cite{DBLP:journals/corr/Sackman15} evenly spreads the keyspace across nodes and achieves a peak-to-average load ratio of 1 while maintaining strict monotonicity, through the use of universal cycles for permutations when assigning keyspace buckets to nodes. However, because those cycles are of length $n!$, this approach appears of limited practical use, and little empirical study has been provided.

AnchorHash~\cite{DBLP:journals/ton/MendelsonVBLKO21} is a more practical solution that also achieves a peak-to-average load ratio of 1 while maintaining strict monotonicity. It however requires a predefined upper bound $a$ on the number of nodes, which has to be set during initialization and cannot be changed later. It works by allocating an array of $a$ buckets, and associating $n$ of them to the nodes. To map a key to a node, AnchorHash rehashes the key until it hits a bucket associated to a node. The evaluation time of the minimal-memory implementation is $O\left(\left( 1 + \log\left(a/n\right)\right)^2\right)$ and the state uses $O(a)$ memory, forcing $a$ to be chosen conservatively and preventing an arbitrarily large upper bound on $n$.

DxHash~\cite{DBLP:journals/corr/abs-2107-07930} removes the limitation of having a capped number of nodes by always maintaining the property that $a/4 < n \leq a$. The evaluation time is $O(a/n)$, which is essentially constant due to this property. If the number of nodes violates that constraint, it will scale out or shrink down the number of buckets by a factor of 2. Unfortunately, scaling out breaks the monotonicity property by requiring a large fraction of the keys to be remapped to maintain balance.

Jump Consistent Hash~\cite{DBLP:journals/corr/LampingV14}, hereafter JumpHash, is the first solution to have considered the more restrictive \emph{consistent range-hashing} problem. It restricts the scope of the problem by requiring the nodes, or resources, to be indexed sequentially, so that keys are hashed to a range of integers. This is not actually a restriction when adding resources because new resources can be assigned the next available index, but it prevents arbitrary resource removals, as they may cause gaps in the index sequence. The only resource that can be removed is the last added one, which is always the one with the greatest index. This restriction rules out applications that handle resource removals by reassigning keys to the remaining resources, such as in distributed web caching. There are however many situations that by design do not require the handling of arbitrary resource removals. This is for instance the case when partitioning data in a database into a number of shards, each of which will persist regardless of hardware faults, which are handled by other mechanisms such as redundancy rather than dynamic key reallocation.

JumpHash shows that under these conditions, both strict monotonicity and perfect balance can be achieved in practice. JumpHash works by using a 64-bit linear congruential pseudorandom generator whose seed corresponds to the key being hashed. It follows (or \emph{jumps} across) the sequence of pseudorandom values until the value is greater than $n$ at which point it returns the penultimate value as the hash value. JumpHash has $O(\log n)$ time complexity.

Recently, Coluzzi, Brocco and Leidi~\cite{DBLP:conf/sebd/ColuzziBL23} have surveyed the field of consistent hashing along with some experimental comparisons of the above algorithms. They find AnchorHash to be the most performant consistent hashing function, with DxHash close behind, and JumpHash outperforming them both when restricting deletions. The same authors with Antonucci~\cite{coluzzi2024mementohash} then propose MementoHash, which augments any consistent range-hashing algorithm to allow for arbitrary deletions by incorporating a minimal data structure that tracks all the removed nodes. They report their consistent hashing algorithm (when using JumpHash as the consistent range-hashing algorithm) to have comparable evaluation time to JumpHash, while using considerably less memory than AnchorHash or DxHash.

\subsection{Preliminaries}

Consistent hashing commonly relies on the concept of a \emph{ranged hash function}~\cite{DBLP:conf/stoc/KargerLLPLL97}, which in addition to the key takes as input the set of resources that the hash function may hash the key to. In practice, that subset designates the set of nodes that are active in the pool at some point.

As mentioned previously, we focus on the more particular situation where the resources can be indexed by ranges of the form $[n] = \{0, \dots, n-1\}$ with $n \in \mathbb{N}^*$, the natural numbers without zero. That is, we intend to build a hash function of the form $f: X \times \mathbb{N}^* \to \mathbb{N}$ where $f(x, n) \in [n]$ for all keys $x$ in the keyspace $X$, and all $n \in \mathbb{N}^*$. We expect from this function the properties of \emph{monotonicity} and \emph{regularity} that formalize consistency and balance, and that we define next.

\begin{definition}[monotonicity]
  $f$ is \emph{monotone} if for all $x \in X$ and $n, n' \in \mathbb{N}^*$ with $n < n'$, $f(x, n') < n$ implies $f(x, n') = f(x, n)$.
\end{definition}

Monotonicity is the property that as one changes the hash table, that is to say the set of values that the function may hash to, keys may not be remapped across hash values that were part of the hash table both before and after resizing the hash table.

Before stating the next definitions, we introduce the following notation: for $a, b \in \mathbb{R}$ and $\epsilon \in [0, 1)$, we will write:
\[a \overset{\epsilon}\approx b \textrm{ if } b(1-\epsilon) \leq a \leq b(1+\epsilon).\]

\begin{definition}[$\epsilon$-regularity]
  Let $\epsilon \in [0,1)$. $f$ is \emph{$\epsilon$-regular} if for all $n \in \mathbb{N}^*$ and $z \in [n]$, $$\Pr_{x \in X} \left( f(x, n)=z \right) \overset{\epsilon}\approx \frac{1}{n}.$$
\end{definition}

Regularity formalizes load balance, or the uniform distribution of keys across resources. Given that a key is drawn uniformly at random, it is equally likely to be hashed to any of the hash values as one another. We use $\epsilon$-regularity to sweep aside the case where the number of keys is finite and is not a multiple of the number of distinct hash values by setting $\epsilon \geq 1/|X|$.

We will also use the property of $(\epsilon, k)$-wise independence of a family of hash functions of the form $h: \Sigma \to \mathcal{R}$, which we define as follows.

\begin{definition}[$(\epsilon, k)$-wise independence]
  Let $\epsilon \in [0, 1)$ and $k \in \mathbb{N}$. A hash family $\mathcal{H} = \left\{ h: \Sigma \to \mathcal{R} \right\}$ is \emph{$(\epsilon, k)$-wise independent} if $\mathcal{R}$ is finite and, for all pairwise distinct $\sigma_1,\dots,\sigma_k \in \Sigma$ and for all $z_1,\dots,z_k \in \mathcal{R}$,
  $$
    \Pr_{h \in \mathcal{H}} \left( \bigwedge_{i=1}^{k} h(\sigma_i) = z_i \right)
    \overset{\epsilon}\approx \frac{1}{|\mathcal{R}|^k}.
  $$
\end{definition}

Observe that $(\epsilon, k)$-wise independence implies $(\epsilon, l)$-wise independence for all integers $l \leq k$.

\section{FlipHash}

We now describe our algorithm. FlipHash is a ranged hash function $f_{\mathcal{H}, \sigma} \colon X \times \{1, \dots, n\} \to [n]$. FlipHash builds upon any family of hash functions $\mathcal{H} = \left\{ h_x: \Sigma \to [2^q] \right\}_{x \in X}$, with $q \in \mathbb{N}^*$, and $q \geq \log_2 n$. For concreteness, we can imagine the keyspace $X$ being strings, $\Sigma = [2^{32}]$ the 32-bit integers, and $q=64$. It also assumes the existence of a way to build unique elements in $\Sigma$ from pairs of integers, specifically, of an injective function $\sigma \colon \{0, \dots, q-1\} \times \{0, \dots, m\} \to \Sigma$ for some integer $m$. We will show that under the hypothesis that $\mathcal{H}$ is $(\epsilon, k)$-wise independent, $f_{\mathcal{H}, \sigma}$ is both monotone and $O(\epsilon + \frac{1}{2^k})$-regular, hence providing stability and load balance. Section~\ref{implementation} discusses the choice of the hash family $\mathcal{H}$ and how $\sigma$ is built in practice.

We will start by describing how FlipHash hashes a key $x \in X$ when the number of resources is a power of 2. We call $\tilde{f}_{\mathcal{H}, \sigma}$ the restriction of $f_{\mathcal{H}, \sigma}$ to the ranges of the form $[2^r]$ for integers $r \leq q$. We will then generalize to any range $[n]$ with $n \leq 2^q$ by building $f_{\mathcal{H}, \sigma}$ from $\tilde{f}_{\mathcal{H}, \sigma}$. For simplicity, we will omit the subscript in the notation from now on and we will simply write $\tilde{f}$ and $f$ respectively.

\subsection{Case where the number of resources $n$ is a power of 2}

\begin{algorithm}[ht]
  \DontPrintSemicolon
  \caption{$\tilde{f}$ (FlipHash, $n = 2^r$)}
  \BlankLine
  \KwIn{key $x \in X$, $r \in \{0, \dots, q\}$}
  \KwOut{$\tilde{f}(x, 2^r) \in [2^r]$}
  \BlankLine
  $a \leftarrow h_x(\sigma(0,0)) \bmod 2^r$\tcp*[l]{lowest $r$ bits}
  $b \leftarrow  \lfloor \log_2 a \rfloor \textrm{ if } a > 0 \textrm{ else } 0$\;
  $c \leftarrow h_x(\sigma(b, 0)) \bmod 2^b$\tcp*[l]{lowest $b$ bits}
  \Return $a \oplus c$\;
  \label{alg:pow2}
\end{algorithm}

Given a number of resources $n = 2^r$ for some integer $r \leq q$, Algorithm~\ref*{alg:pow2} shows how a key $x \in X$ is hashed into one of the resources: $\tilde{f}(x, 2^r) \in [2^r]$. It first hashes the key $x$ and $\sigma(0,0)$ using the underlying hash family $\mathcal{H}$ and keeps the $r$ least significant bits in $a$. $b$ records the position of the most significant bit of $a$ that is equal to 1. $x$ is then hashed again with $\sigma(b,0)$ to generate a second hash, of which the $b$ least significant bits are kept in $c$. Finally $c$ is used to flip some of the bits of the first hash $a$. This last operation, essential to the properties of FlipHash as we will show below, is what inspired the name of our algorithm.

Given some $x$, $\mathcal{H}$, and $\sigma$, that we assume produce specific values of $h_x(\sigma(b,0))$, we illustrate the behavior of Algorithm~\ref*{alg:pow2} as we increase $r$ in Table~\ref*{table:algpow2ex}. As an example, consider the case where the key $x$ is first hashed into $h_x(\sigma(0,0)) = 1011_2$, and where $r = 2$. The value of $\sigma(0,0)$ itself does not matter at this point. $a$ is built from the 2 least significant bits of the hash: $a = 0011_2$, and $b = 1$. $c$ contains the single least significant bits of the second hash, which could be $c = 0001_2$, and flips some of the bits of $a$ to generate $\tilde{f}(x, 2^2) = 0011_2 \oplus 0001_2 = 0010_2 = 2$. As we double the number of resources by incrementing $r$ from $2$ to $3$, it happens that the additional bit of $h_x(\sigma(0,0))$ that we are using to build $a$ is $0$. $a$ therefore stays the same, as do $b$, $c$, and the final flipped hash value $\tilde{f}(x, 2^3) = 0010_2 = 2$. However, incrementing $r$ once more from $3$ to $4$ changes the value of $a$ from $0011_2$ to $1011_2$. It follows that $b = 3$ and the second hash differs from the previous instance: $c$ could now be $0101_2$. Finally, the key $x$ is hashed to a new value $\tilde{f}(x, 2^4) = 1011_2 \oplus 0101_2 = 1110_2 = 14 \geq 2^3$.

\begin{table}[h]
  \centering
  \begin{tabular}{|c c c c c c|}
    \hline
    $r$   & $a$      & $b$ & $h_x(\sigma(b, 0))$ & $c$      & $\tilde{f}(x, 2^r)$ \\ [0.5ex]
    \hline
    $0$   & $0000_2$ & $0$ & $1011_2$            & $0000_2$ & $0000_2 = 0$        \\
    $1$   & $0001_2$ & $0$ & $1011_2$            & $0000_2$ & $0001_2 = 1$        \\
    $2,3$ & $0011_2$ & $1$ & $0101_2$            & $0001_2$ & $0010_2=2$          \\
    $4$   & $1011_2$ & $3$ & $1101_2$            & $0101_2$ & $1110_2=14$         \\[1ex]
    \hline
  \end{tabular}
  \caption{Values taken by the variables of Algorithm~\ref*{alg:pow2} for $q=4$, assuming that $h_x(\sigma(0,0)) = 11 = 1011_2$, $h_x(\sigma(1,0)) = 5 = 0101_2$, and $h_x(\sigma(3,0)) = 13 = 1101_2$.}
  \label{table:algpow2ex}
\end{table}

In this particular instance, it is to be noted that as the number of resources is gradually increased, the hash stays the same ($\tilde{f}(x, 2^3) = \tilde{f}(x, 2^2)$) or is updated to map to one of the newly added resources ($\tilde{f}(x, 2^4) \geq 2^3$). This is an illustration of the monotonicity of $\tilde{f}$, which Theorem~\ref{th:pow2monotonicity} states to be generally ensured.

\begin{theorem}\label{th:pow2monotonicity}
  $\tilde{f}$ is monotone.
\end{theorem}

\begin{proof}
  It is enough to show that, for all $r \in \{0, \dots, q\}$ and $x \in X$, incrementing $r$ by one either leaves the hash unchanged or updates it to a value that could not be reached before, i.e., greater than or equal to $2^r$.

  Incrementing $r$ by one may only affect $a$ by updating its $(r+1)$-th least significant bit from 0 to either 0 or 1. If the bit stays 0, $a$, $b$ and $c$ are all unchanged, and $\tilde{f}(x, 2^{r+1}) = \tilde{f}(x, 2^r)$. Otherwise, the bit is changed to 1, $b = r$, and the $(r+1)$-th least significant bit of $a \oplus c$ is 1, thus $\tilde{f}(x, 2^{r+1}) \geq 2^r$.
\end{proof}

Next we show the regularity of $\tilde{f}$, i.e., $\tilde{f}(x, 2^r)$ is uniform over $[2^r]$ given that the key $x$ is picked uniformly at random. Requiring $h_x(\sigma(0,0))$ to be uniform given that $x$ is picked at random is not enough because the second hash used for $c$ could depend on the first hash used for $a$ in a way that favors some values of $a \oplus c$ over others (consider the extreme case where every $h_x$ is a constant function independent of $\sigma$, which would cause the last significant bits of $a \oplus c$ to always be $0$). Therefore, we also require independence between $h_x(\sigma(0,0))$ and $h_x(\sigma(b,0))$, which can be ensured with the pairwise independence of the hash family $\mathcal{H}$. This is formalized later in Corollary~\ref{corollary:pow2reg}.

When generalizing the approach to any value of $n$ rather than only powers of 2, we will actually need more than the regularity of $\tilde{f}$, and we will leverage Lemma~\ref{lemma:pow2ind}. It informally states that the increasing hash values produced by $\tilde{f}$, as the range of possible values $[2^r]$ grows, along with those produced by $\mathcal{H}$, across distinct seeds, are all mutually independent. In particular, that implies that when increasing the number of resources, the keys that were mapped to a single resource and get remapped to new resources are spread across all the new resources, which makes upscaling resources likely to spread hotspots across a large number of resources, rather than only moving them to single new resources.

This last property is the motivation for the flip operation of the algorithm, i.e., for returning $a \oplus c$ rather than $a$ directly. Returning $a$ would also achieve both the monotonicity and the regularity of $\tilde f$, but it would not ensure that as $r$ increases, keys are remapped to new resources in a way that is independent of the previous hash values, because the lowest bits of the hash would stay the same.

\begin{lemma}\label{lemma:pow2ind}
  Let $\epsilon \in [0,1)$, $k_{\tilde f} \in \mathbb{N}$, $k_h \in \mathbb{N}$ and let $\mathcal{H}$ be $(\epsilon, k_{\tilde f}+k_h+1)$-wise independent. Let $r_1, \dots, r_{k_{\tilde f}} \in \{0, \dots, q\}$, and $z_1, \dots, z_{k_{\tilde f}} \in [2^q]$ such that:
  \[z_1 < 2^{r_1} \leq z_2 < \dots < 2^{r_{k_{\tilde f}}}.\]

  Let $\rho_1, \dots, \rho_{k_h} \in \Sigma \setminus \{ \sigma(b, 0) \colon b \in \{0, \dots, q-1\} \}$ distinct, and $\eta_1, \dots, \eta_{k_h} \in [2^q]$. Then:
  \[
    \Pr_{x \in X} \left( \bigwedge_{i=1}^{k_{\tilde f}} \tilde{f}(x, 2^{r_i}) = z_i \land \bigwedge_{i=1}^{k_h} h_x(\rho_i) = \eta_i \right)
    \overset{\epsilon}\approx
    2^{-\left(\sum_{i=1}^{k_{\tilde f}} r_i + q k_h\right)}.
  \]
\end{lemma}

\begin{proof}
  Let $P$ be the probability of the left-hand side of the equality and let $l \colon a \mapsto \lfloor \log_2 a \rfloor \textrm{ if } a > 0 \textrm{ else } 0$. Because $c$ only flips the bits after the most significant 1 bit of $a$, $l(a \oplus c) = l(a)$. Similarly, for all integers $\zeta$, $l( z_i \oplus ( \zeta \bmod 2^{l(z_i)} ) ) = l(z_i)$. Thus, for $i \in \{1, \dots, k_{\tilde f}\}$, we can replace $l(a)$ with $l(z_i)$, and we have that $\tilde{f}(x, 2^{r_i}) = z_i$ if and only if:
  \begin{align*}
    \left( h_x(\sigma(0,0)) \bmod 2^{r_i} \right) \oplus \left( h_x(\sigma(l(z_i), 0)) \bmod 2^{l(z_i)} \right) = z_i.
  \end{align*}
  It follows that we can express $P$ as a sum of probabilities:
  \begin{align*}
    P = \sum_{(\theta_0, \dots, \theta_{k_{\tilde f}}) \in \Theta} \Pr_{x \in X} \left( h_x(\sigma(0,0)) = \theta_0 \land \bigwedge_{i=1}^{k_{\tilde f}} h_x(\sigma(l(z_i),0)) = \theta_i \land \bigwedge_{i=1}^{k_h} h_x(\rho_i) = \eta_i  \right).
  \end{align*}
  with:
  \begin{align*}
    \Theta = \left\lbrace (\theta_0, \dots, \theta_{k_{\tilde f}}) \in [2^q]^{k_{\tilde f}+1} \colon \forall i \in \{1, \dots, k_{\tilde f}\}, \left( \theta_0 \bmod 2^{r_i} \right) \oplus \left( \theta_i \bmod 2^{l(z_i)} \right) = z_i \right\rbrace.
  \end{align*}

  $l(z_i)$ is equal to zero if and only if $z_i = 0$ or $1$, which may happen if $i = 1$ or $2$. Let $J = \{i \in \{1, \dots, k_{\tilde f}\} \colon l(z_i) = 0\}$. Observing that for all $i \in J$, the inner conjunction implies $\theta_i = \theta_0$, we can rewrite $P$ as:
  \begin{align*}
    P = \sum_{(\theta_i) \in \Theta \cap \Lambda_J} \Pr_{x \in X} \left( h_x(\sigma(0,0)) = \theta_0 \land \bigwedge_{\substack{i=1 \\ l(z_i) \neq 0}}^{k_{\tilde f}} h_x(\sigma(l(z_i),0)) = \theta_i \land \bigwedge_{i=1}^{k_h} h_x(\rho_i) = \eta_i \right).
  \end{align*}
  with $\Lambda_J = \left\lbrace (\theta_0, \dots, \theta_{k_{\tilde f}}) \in [2^q]^{k_{\tilde f}+1} \colon \forall i \in J, \theta_i = \theta_0 \right\rbrace$.
  Each term is the probability of the conjunction of at most $k_{\tilde f}+k_h+1$ events of the form $h_x(\sigma_j) = \mu_j$ with pairwise distinct $\sigma_j$. Because $\mathcal{H}$ is $(\epsilon, k_{\tilde f}+k_h+1)$-wise independent, we have that:
  $$
    P \overset{\epsilon}\approx \frac{|\Theta \cap \Lambda_J|}{2^{q(1+k_{\tilde{f}}-|J|+k_h)}}.
  $$
  It remains to calculate $|\Theta \cap \Lambda_J|$. For all $i$, $\left( \theta_0 \bmod 2^{r_i} \right) \oplus \left( \theta_i \bmod 2^{l(z_i)} \right) = z_i$ implies:
  $$
    l \left(\theta_0 \bmod 2^{r_i}\right) = l\left(z_i \oplus \left( \theta_i \bmod 2^{l(z_i)} \right)\right) = l(z_i),
  $$
  i.e., any bit of $\theta_0$ of index between $l(z_i)$ (excluded) and $r_i$ (included) is equal to zero. Given that $l(z_1) < r_1 \leq l(z_2) < \dots < r_{k_{\tilde f}}$ by hypothesis, there are $2^{q - \sum_{i=1}^{k_{\tilde{f}}}(r_i - l(z_i))}$ values of $\theta_0$ in $[2^q]$ that have this property for all $i$. For all such $\theta_0$, and for all $i$, there are $2^{q-l(z_i)}$ values of $\theta_i$ in $[2^q]$ such that $\left( \theta_0 \bmod 2^{r_i} \right) \oplus \left( \theta_i \bmod 2^{l(z_i)} \right) = z_i$. All things considered:
  \begin{align*}
    |\Theta \cap \Lambda_J|
     & = 2^{q - \sum_{i=1}^{k_{\tilde{f}}}(r_i - l(z_i))} \prod_{\substack{i=1 \\ i \notin J}}^{k_{\tilde{f}}} 2^{q-l(z_i)} \\
     & = 2^{q(1+k_{\tilde f}-|J|) - \sum_{i=1}^{k_{\tilde{f}}}r_i}.
  \end{align*}
  And $P$ reduces to:
  $$
    P \overset{\epsilon}\approx 2^{-\left(\sum_{i=1}^{k_{\tilde f}}r_i + q k_h\right)}.
  $$
\end{proof}

In particular, from the lemma with $k_{\tilde{f}} = 1$ and $k_h = 0$ follows the previously mentioned result about the regularity of $\tilde{f}$, formalized in Corollary~\ref{corollary:pow2reg}.

\begin{corollary}\label{corollary:pow2reg}
  Let $\epsilon \in [0,1)$ and let $\mathcal{H} $ be $\epsilon$-pairwise independent. Then, $\tilde{f}$ is $\epsilon$-regular.
\end{corollary}

\subsection{Generalization to any number of resources $n \leq 2^q$}

\begin{algorithm*}[ht]
  \DontPrintSemicolon
  \caption{$f$ (FlipHash)}
  \BlankLine
  \KwIn{key $x \in X$, number of resources $n \in \{1, \dots, 2^q\}$}
  \KwOut{$f(x, n) \in [n]$}
  \BlankLine
  $r \leftarrow \lceil \log_2 n \rceil$\;
  $d \leftarrow \tilde{f}(x, 2^r)$\;
  \lIf{$d < n$}{
    \Return $d$\tcp*[f]{(A)}
  } \Else {
    \For{$i = 1 \dots m$} {
      $e_i \leftarrow h_x(\sigma(r-1, i)) \bmod 2^r$\tcp*[l]{lowest $r$ bits}
      \lIf{$e_i < 2^{r-1}$}{
        \Return $\tilde{f}(x, 2^{r-1})$\tcp*[f]{(B)}
      } \lElseIf{$e_i < n$}{
        \Return $e_i$\tcp*[f]{(C)}
      }
    }
    \Return{$\tilde{f}(x, 2^{r-1})$}\tcp*[f]{(D)\kern 0.3em}
  }
  \label{alg:fliphash}
\end{algorithm*}

Algorithm~\ref{alg:fliphash} generalizes Algorithm~\ref{alg:pow2} to any number of resources $n$. It leverages Algorithm~\ref{alg:pow2} using the power of 2 immediately greater than or equal to $n$, $2^r$. If the value that Algorithm~\ref{alg:pow2} returns is within the accepted range $[n]$, it is returned with the return statement (A). Otherwise, the algorithm repetitively hashes the key $x$ with varying values $\sigma(r-1,i)$ until one of the hashes, masked to the least $r$ significant bits, is within the range $[n]$. At this point, it either returns it with (C) if it is greater than or equal to $2^{r-1}$, or return with (B) the value that Algorithm~\ref{alg:pow2} generates with the previous power of 2. The latter case ensures monotonicity. The number of iterations is also bounded by $m$ to ensure that the algorithm terminates, without any assumptions on $\mathcal{H}$.

Given some $x$, $\mathcal{H}$, and $\sigma$, that we assume produce the same specific values of $h_x(\sigma(b,i))$ as the ones of Table~\ref*{table:algpow2ex}, Table~\ref*{table:algex} illustrates the behavior of Algorithm~\ref{alg:fliphash} for successive values of $n$.

\begin{table}[h]
  \centering
  \begin{tabular}{|c c c c c c c c|}
    \hline
    $n$       & $r$ & $d$  & $e_1$ & $e_2$ & $e_3$ & $e_4$ & $f(x,n)$              \\ [0.5ex]
    \hline
    $1$       & $0$ & $0$  &       &       &       &       & $d=0$                 \\
    $2$       & $1$ & $1$  &       &       &       &       & $d=1$                 \\
    $3,4$     & $2$ & $2$  &       &       &       &       & $d=2$                 \\
    $5,6,7,8$ & $3$ & $2$  &       &       &       &       & $d=2$                 \\
    $9,10,11$ & $4$ & $14$ & $12$  & $11$  & $15$  & $6$   & $\tilde{f}(x, 2^3)=2$ \\
    $12$      & $4$ & $14$ & $12$  & $11$  &       &       & $e_2=11$              \\
    $13, 14$  & $4$ & $14$ & $12$  &       &       &       & $e_1=12$              \\
    $15,16$   & $4$ & $14$ &       &       &       &       & $d=14$                \\
    \hline
  \end{tabular}
  \caption{Values taken by the variables of Algorithm~\ref*{alg:fliphash} for $q = 4$, assuming that $h_x(\sigma(0,0)) = 11$, $h_x(\sigma(1,0)) = 5$, $h_x(\sigma(3,0)) = 13$, $h_x(\sigma(3,1)) = 12$, $h_x(\sigma(3,2)) = 11$, $h_x(\sigma(3,3)) = 15$, and $h_x(\sigma(3,4)) = 6$.}
  \label{table:algex}
\end{table}

Algorithm~\ref{alg:fliphash} builds on the monotonicity of Algorithm~\ref{alg:pow2} and itself ensures monotonicity, as shown by Theorem~\ref{th:powreg}.

\begin{theorem}\label{th:powreg}
  FlipHash is monotone.
\end{theorem}

\begin{proof}
  It is enough to show that, for all $x \in X$ and $n \in \{1, \dots, 2^q\}$, $f(x, n+1) = f(x, n)$ or $f(x, n+1) = n$.
  By construction, $n \leq 2^r$.
  \begin{itemize}
    \item  If $n < 2^r$, incrementing $n$ leaves both $r$ and $d$ unchanged: $r' = r$ and $d' = d$, where primed variables denote the values of the variables when the algorithm is run with $n' = n+1$ as an input.
          \begin{itemize}
            \item If $d < n'$, then $f(x, n') = d' = d$ by return statement (A).
                  \begin{itemize}
                    \item If $d < n$, then $f(x, n') = d = f(x, n)$.
                    \item Else, $d = n$, and $f(x, n') = d = n$.
                  \end{itemize}
            \item Else, running the algorithm with either $n$ or $n'$ leads to the top-level "else" branch. For convenience, we will define $e_{m+1} = e'_{m+1} = 0$ so that the return statement (D) is equivalent to an additional loop iteration with $i = m+1$. Because $2^{r-1} < n < n'$, the loop iterates until $e_i < n$, respectively, $e'_i < n'$. Let $i_0$ (resp., $i'_0$) be the lowest $i \in \{1, \dots, m+1\}$ such that $e'_i < n$ (resp., $e_i < n'$). Because $r' = r$, we have that $e'_i = e_i$ for all $i \in \{1, \dots, m\}$, and t $i'_0 \leq i_0$.
                  \begin{itemize}
                    \item If $i'_0 = i_0$, either $f(x, n') = \tilde{f}(x, 2^r) = f(x, n)$ by return statement (B), or $f(x, n') = e_{i_0} = f(x, n)$ by return statement (C).
                    \item Else, $i'_0 < i_0$, hence $n \leq e_{i'_0} < n'$ and $e_{i'_0} = n > 2^{r-1}$. It follows that $f(x, n') = e_{i'_0} = n$ by return statement (C).
                  \end{itemize}
          \end{itemize}
    \item Else, $n = 2^r$, and $r' = r+1$. Because $d < 2^r = n$, $f(x, n) = d = \tilde{f}(x, 2^r)$ by return statement (A).
          \begin{itemize}
            \item If $d' = d$, because $d' = d < n < n'$, then by return statement (A), $f(x, n') = d' = d = f(x, n)$.
            \item Else, because $\tilde{f}$ is monotone, $d' \geq 2^r = n$.
                  \begin{itemize}
                    \item If $d' = n$, because $d' < n'$, then by return statement (A), $f(x, n') = d' = n$.
                    \item Else, $d' \geq n'$. With $i'_0$ as defined above,
                          \begin{itemize}
                            \item If $e_{i'_0} < 2^{r'-1}$, then by return statement (B), $f(x, n') = \tilde{f}(x, 2^{r'-1}) = \tilde{f}(x, 2^r) = f(x, n)$.
                            \item Else, because $n = 2^{r'-1} \leq e_{i'_0} < n' = n+1$, $e_{i'_0} = n$. It follows that by return statement (C), $f(x, n') = e_{i'_0} = n$.
                          \end{itemize}
                  \end{itemize}
          \end{itemize}
  \end{itemize}
\end{proof}

We prove the regularity properties of FlipHash in Theorem~\ref{th:fliphashreg}.

\begin{theorem}\label{th:fliphashreg}
  Provided that $\mathcal{H}$ is $(\epsilon, k)$-wise independent with $\epsilon \in [0,1)$ and $k \geq 3$, FlipHash, as defined by Algorithm~\ref{alg:fliphash} with $m = k - 3$, is $O\left( \epsilon + \frac{1}{2^k} \right)$-regular.
\end{theorem}

\begin{proof}
  Let $z \in [n]$. The probability that Algorithm~\ref{alg:fliphash} returns $z$ is the sum of the probabilities that each of its return statements is reached and returns $z$:
  $$
    \Pr_{x \in X} \left( f(x, n) = z \right)
    = A + \sum_{i=1}^m \left( B_i + C_i \right) + D.
  $$
  We will cover each of those terms, starting with $A$. The return statement (A) is reached if and only if $d = \tilde{f}(x, 2^r) < n$, in which case $d$ is returned. Thus:
  \begin{align*}
    A & = \Pr_{x \in X} \left( \tilde{f}(x, 2^r) < n \land \tilde{f}(x, 2^r) = z \right) \\
      & = \Pr_{x \in X} \left( \tilde{f}(x, 2^r) = z \right)                             \\
      & \overset{\epsilon}\approx \frac{1}{2^r}.
  \end{align*}
  where the second relation follows from the hypothesis $z < n$, and the last relation follows from the $\epsilon$-regularity of $\tilde{f}$ given Corollary~\ref{corollary:pow2reg}.

  The return statement (B) only produces values in $[2^{r-1}]$. Therefore, if $z \geq 2^{r-1}$, then for all $i \in \{1, \dots, m\}$, $B_i = 0$. Otherwise, $z < 2^{r-1}$, and $B_i$ can be expressed from the successive predicates that need to evaluate to true to lead to the return statement (B) at the $i$-th iteration:
  \begin{align*}
    B_i = \Pr_{x \in X} \left( \tilde{f}(x, 2^r) \geq n \land \bigwedge_{j=1}^{i-1} e_j(x) \geq n \land e_i(x) < 2^{r-1} \land \tilde{f}(x, 2^{r-1}) = z \right),
  \end{align*}
  where for $j \in \{1, \dots, m\}$, $e_j(x) = h_x(\sigma(r-1, j)) \bmod 2^r$. Because $\sigma$ is injective, Lemma~\ref{lemma:pow2ind} with $k_{\tilde{f}} = 2$ and $k_h = i \leq m$, together with the law of total probability, allow rewriting $B_i$ as the product of the probabilities of the individual events:
  $$
    B_i \overset{\epsilon}\approx
    \frac{2^r-n}{2^r}
    \left( \prod_{j=1}^{i-1} \frac{2^r-n}{2^r} \right)
    \frac{2^{r-1}}{2^r}
    \frac{1}{2^{r-1}}
    = \frac{1}{2^r}
    \left( 1 - \frac{n}{2^r} \right)^i.
  $$

  The same approach is taken for the return statement (C), which does not produce values in $[2^{r-1}]$. Therefore, if $z < 2^{r-1}$, then for all $i \in \{1, \dots, m\}$, $C_i = 0$. Otherwise, $z \geq 2^{r-1}$, and $C_i$ can be written as:
  \begin{align*}
    C_i = \Pr_{x \in X} \left( \tilde{f}(x, 2^r) \geq n \land \bigwedge_{j=1}^{i-1} e_j(x) \geq n \land 2^{r-1} < e_i(x) \leq n \land e_i(x) = z \right).
  \end{align*}
  Likewise, Lemma~\ref{lemma:pow2ind} with $k_{\tilde{f}} = 1$ and $k_h = i \leq m$ gives:
  $$
    C_i \overset{\epsilon}\approx
    \frac{2^r-n}{2^r}
    \left( \prod_{j=1}^{i-1} \frac{2^r-n}{2^r} \right)
    \frac{1}{2^r}
    = \frac{1}{2^r}
    \left( 1 - \frac{n}{2^r} \right)^i.
  $$
  Observing that this is the same expression as $B_i$ when $z < 2^{r-1}$, the sum of $B_i$ and $C_i$ forms a geometric series. Using the closed-form formula for the sum of its first terms, we have that:
  \begin{align*}
    \sum_{i=1}^m \left( B_i + C_i \right)
     & \overset{\epsilon}\approx
    \frac{1}{2^r}
    \sum_{i=1}^m \left( 1 - \frac{n}{2^r} \right)^i                                                     \\
     & = \left( \frac{1}{n} - \frac{1}{2^r} \right) \left( 1 - \left( 1-\frac{n}{2^r} \right)^m \right) \\
     & = \left( \frac{1}{n} - \frac{1}{2^r} \right) \left( 1 + O\left(\frac{1}{2^m}\right) \right),
  \end{align*}
  where the last relation uses the facts that $2^{r-1} < n \leq 2^r$.

  Finally, $D$ can be written as:
  \begin{align*}
    D = \Pr_{x \in X} \left( \tilde{f}(x, 2^r) \geq n \land \bigwedge_{i=1}^m e_i(x) \geq n \land \tilde{f}(x, 2^{r-1}) = z \right).
  \end{align*}
  Lemma~\ref{lemma:pow2ind} with $k_{\tilde{f}} = 1$ and $k_h = m$ gives:
  \begin{align*}
    D & \overset{\epsilon}\approx
    \left( \frac{2^r-n}{2^r} \right)^m \frac{1}{2^{r-1}}     \\
      & = \frac{1}{2^{r-1}} \left( 1-\frac{n}{2^r} \right)^m \\
      & = \frac{1}{n} O\left( \frac{1}{2^m} \right).
  \end{align*}

  We can finally combine the probabilities matching every return statement:
  \begin{align*}
    \Pr_{x \in X} \left( f(x, n) = z \right)
     & \overset{\epsilon}\approx
    \frac{1}{2^r} + \left( \frac{1}{n} - \frac{1}{2^r} \right) \left( 1 + O\left(\frac{1}{2^m}\right) \right) + \frac{1}{n} O\left( \frac{1}{2^m} \right) \\
     & = \frac{1}{n} \left( 1 + O\left( \frac{1}{2^m} \right) \right).
  \end{align*}
  By definition of $\overset{\epsilon}\approx$, and because $k = m + 3$, we prove the $O\left( \epsilon + \frac{1}{2^k} \right)$-regularity of $f$:
  \begin{align*}
    \Pr_{x \in X} \left( f(x, n) = z \right)
     & = \frac{1}{n}
    \left( 1 + O\left( \frac{1}{2^k} \right) \right)
    \left( 1 + O\left( \epsilon \right) \right) \\
     & = \frac{1}{n}
    \left( 1 + O\left( \epsilon + \frac{1}{2^k} \right)\right).
  \end{align*}
\end{proof}

Finally, provided that computing $\sigma$ is a constant-time operation, we state in Theorem~\ref{th:fliphashtime} that the time complexity of FlipHash is independent of $n$.

\begin{theorem}\label{th:fliphashtime}
  Let $c$ be the time complexity of $\mathcal{H}$. Provided that $\mathcal{H}$ is $(\epsilon, k)$-wise independent with $\epsilon \in [0,1)$ and $k \in \mathbb{N}$, the time complexity of Algorithm~\ref{alg:fliphash} with $m \leq k$ is $O(c)$ on average and $O(mc)$ at worst.
\end{theorem}

\begin{proof}
  The worst-case complexity results from the fact that Algorithm~\ref{alg:fliphash} runs $m$ for-loop iterations at most. The probability that it runs exactly $i$ iterations is upper-bounded by the probability that the first $i-1$ iterations each generate a value $e_j \in [2^r]$ such that $e_j \geq n$, itself upper-bounded by $(1+\epsilon)\prod_{j=1}^{i-1}\frac{2^{r-1}}{2^r} = \frac{1+\epsilon}{2^{i-1}}$ given that $\mathcal{H}$ is $(\epsilon, k)$-wise independent and $n > 2^{r-1}$. The average time complexity is therefore $O(c(1+\epsilon)\sum_{i=1}^m \frac{i}{2^{i-1}}) = O(c)$.
\end{proof}

\section{Evaluation}

\subsection{Implementation}\label{implementation}

Implementing FlipHash requires the choice of a hash algorithm $\mathcal{H}$ that maps pairs of the keys in $X$ of the desired type and the elements of $\Sigma$, to integer values of $[2^q]$ (e.g., the 64-bit integer values $[2^{64}]$). The behavior and the performance of FlipHash depend on those of $\mathcal{H}$. Therefore we strive to use a fast hash algorithm that evenly distributes keys in $X$ over its hash table $[2^q]$, and is seeded by elements in $\Sigma$ across which it shows good hash mutual independence. Because of its computational performance, XXH3~\cite{XXH3} is a good candidate. We were able to empirically validate with statistical tests the regularity of FlipHash when using it, with $m = 64$.

In addition, we built a standalone implementation that, like JumpHash, takes 64-bit integer keys as an input. It uses a custom hash algorithm $\mathcal{H}$ that takes inspiration from bit-mixing constructs~\cite{BitMixer}~\cite{Moremur}, often used in finalizing steps of hash algorithms to generate random-looking uniform values. This enables a simple and fast implementation of FlipHash that also empirically shows the expected regularity. This version can also be used to consistently hash keys of any type by taking as an input the integer output of any hash algorithm, including XXH3. This method has the benefit of hashing the initial input only once and offers a performance gain if the input itself is large. It however has the disadvantage that it loses the ability to use the full entropy of the input when generating the multiple hashes that FlipHash uses and expects to be mutually independent. Instead, it has to rely on the 64 bits of the input key only, similar to JumpHash. In theory, this can degrade the regularity of FlipHash, especially if the range of hash values is large. However, we could not observe such degradation for practical ranges, and we claim that this standalone implementation of FlipHash is suitable for the majority of use cases.

It remains to discuss the construction of the injective function $\sigma \colon \{0, \dots, q-1\} \times \{0, \dots, m\} \to \Sigma$. Hash algorithms are commonly seeded with integers that have at least 32 bits, and both $q$ and $m$ are less than $2^{16}$ for practical purposes. We can therefore use $\sigma \colon (r,i) \mapsto r+i2^{16}$. In addition, we can use $\sigma$ to make FlipHash itself seeded, by using $\sigma_s \colon (r,i) \mapsto (r+i2^{16}) \oplus s$, where $s$ is the seed.

We have made the standalone implementation of FlipHash in Rust available online\footnote{\url{https://github.com/datadog/fliphash-rs}}, as well as the one that leverages XXH3. This also includes testing and benchmarking code.

\subsection{Performance}

\begin{figure}[ht]
  \centering
  \includegraphics[width=0.8\linewidth]{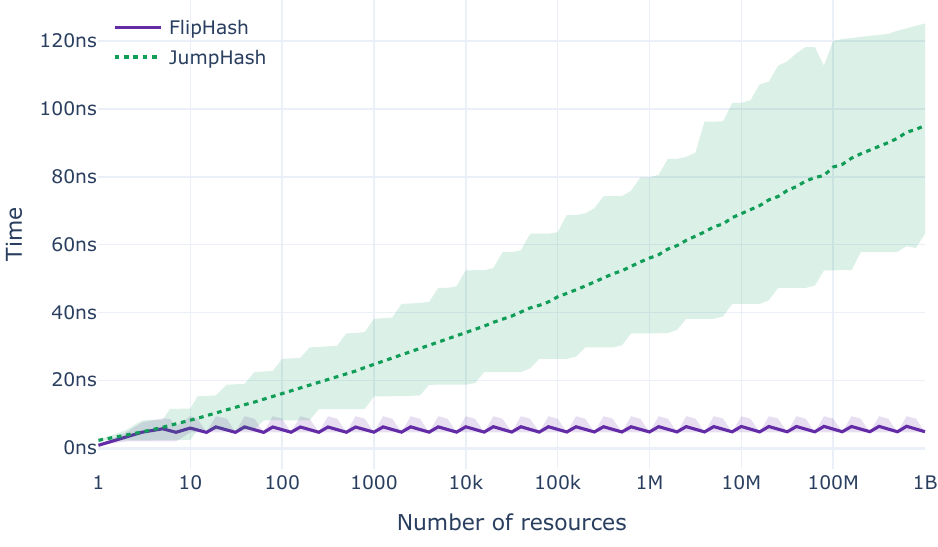}
  \caption{Average (line) and interdecile (filled area) evaluation wall times of FlipHash and JumpHash.}
  \label{fig:time}
\end{figure}

The following benchmarks have been run on an Intel\textsuperscript{\textregistered} Xeon\textsuperscript{\textregistered} Platinum 8375C CPU @ $\qty{2.9}{GHz}$, using the standalone version of FlipHash, which like JumpHash, takes 64-bit integer keys as input. Figure~\ref{fig:time} shows the evaluation times of FlipHash and JumpHash for numbers of resources between one and one billion. In addition to the average evaluation time across keys, it shows the interdecile range, i.e., the range between the 10\textsuperscript{th} and the 90\textsuperscript{th} percentiles. The interdecile range is relevant as a measure of the evaluation time dispersion, given that the numbers of iterations that both FlipHash and JumpHash run in their internal loops vary across keys. The graph shows the constant asymptotic dependency on the number of resources of the evaluation time for FlipHash, and the logarithmic one for JumpHash. FlipHash takes a few nanoseconds to hash a key on average, however large the number of resources, whereas JumpHash while similarly fast for small numbers of resources, requires tens of nanoseconds for larger values.

\begin{figure}[ht]
  \centering
  \includegraphics[width=0.8\linewidth]{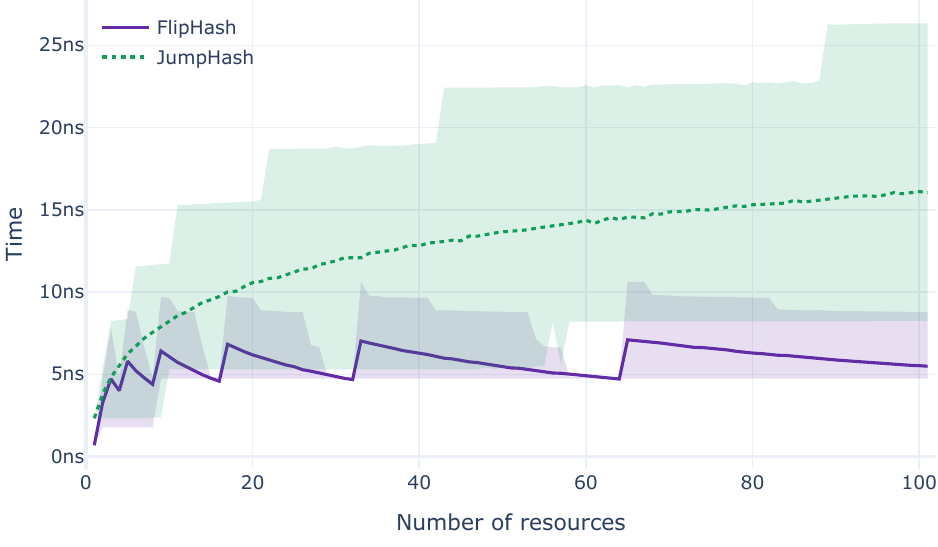}
  \caption{Average (line) and interdecile (filled area) evaluation wall times of FlipHash and JumpHash for small numbers of resources.}
  \label{fig:timesmallrange}
\end{figure}

Figure~\ref{fig:timesmallrange} focuses on hashing with fewer than 100 resources. It shows that FlipHash is substantially faster than JumpHash when hashing with more than 10 resources, which should be the case in the vast majority of applications. The graph also exhibits a characteristic "sawtooth" behavior of the evaluation time of FlipHash. This is because the farther away $n$ is from the next power of 2, namely $2^r$, the more loop iterations FlipHash runs on average, given that both predicates $d<n$ and $e_i<n$ in Algorithm~\ref{alg:fliphash} are less likely to evaluate to true.

\begin{figure}[ht]
  \centering
  \includegraphics[width=0.8\linewidth]{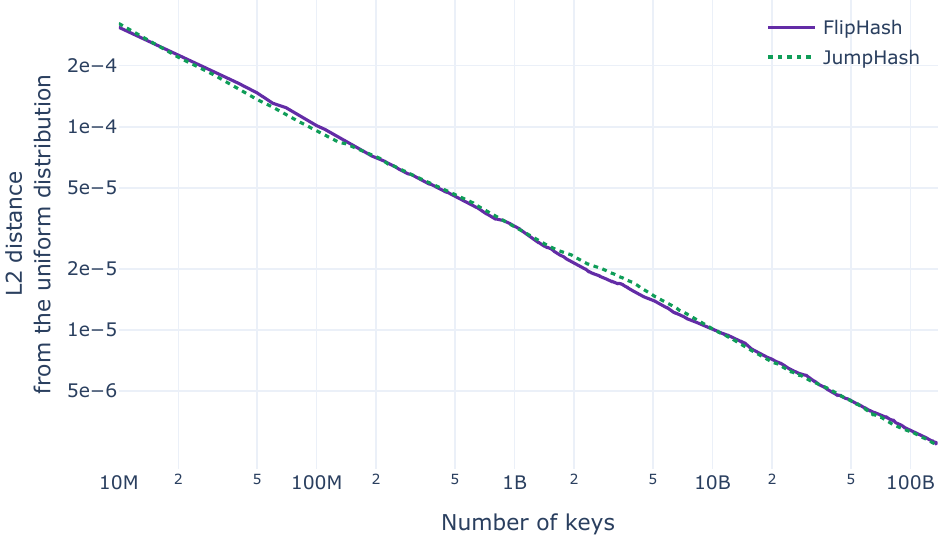}
  \caption{Measure of the uniformity of the key distribution across 1000 resources of FlipHash and JumpHash, with randomly generated keys.}
  \label{fig:regularity}
\end{figure}

In addition to computational performance, we have studied the quality of the output of FlipHash. We have validated, through the use of Chi-Squared statistical tests not only the regularity of FlipHash, but also its ability to generate mutually independent hashes when varying the seed, and mutually independent hashes when varying the number of resources, given that the hashes are pairwise distinct. The latter mutual independence implies that as resources are added, keys are remapped evenly across the newly added resources, which as previously alluded to enables upscales to spread possible hotspots across multiple of the added resources. We conjecture that both those mutual independence properties, across seeds and across numbers of resources, can be formally proven using similar reasoning as Theorem~\ref{th:fliphashreg}, and Lemma~\ref{lemma:pow2ind} on the final version of FlipHash. Furthermore, through the study of the p-values of those statistical tests and the distribution of the keys across the possible hash values, we were not able to exhibit any difference in the regularity and mutual independence behaviors of FlipHash and JumpHash, in either's favor. As an illustration, Figure~\ref{fig:regularity} shows the comparable regularity of FlipHash and JumpHash by hashing randomly generated keys and measuring the L2 distance between the resulting distribution of hashes and the uniform distribution.

\subsection{Comparison with other consistent hashing algorithms}

We have compared FlipHash against JumpHash, because they both are solutions of the consistent range-hashing problem that do not impose additional constraints or requirements. As the solutions of a more general problem, consistent hashing algorithms can also be used for consistent range-hashing. While those solutions support arbitrary resource removals, they in return give up on at least one of the constraints or properties that FlipHash fulfills. One of them is low memory usage. Because those more general solutions need to keep track of the resources that the keys can be hashed to, and because they often use additional data structures to speed up the hashing operation, contrary to FlipHash and JumpHash, none of those alternative consistent hashing algorithms achieve $O(1)$ memory usage~\cite{DBLP:conf/sebd/ColuzziBL23}. The data structures that those algorithms use also need to be initialized and updated as the number of resources varies over time, sometimes at a high computational cost, whereas FlipHash and JumpHash do not require performing such operations.

Furthermore, those more general solutions often break or degrade other properties. Hash-ring consistent hashing~\cite{DBLP:conf/stoc/KargerLLPLL97} does not achieve a peak-to-average ratio equal to 1, effectively creating load imbalance that translates into the need for overprovisioning resources in practice. Maglev~\cite{DBLP:conf/nsdi/EisenbudYCSKMCC16} does not realize strict monotonicity, causing extraneous key reshuffle that some applications cannot tolerate or that need to be handled by implementing specific mechanisms. Maglev also computes expensive initialization and updates. AnchorHash~\cite{DBLP:journals/ton/MendelsonVBLKO21} prevents scalability by requiring a predefined upper bound on the number of resources that has to be set at initialization, and cannot be chosen arbitrarily large because both the memory usage and the computational cost of AnchorHash depend on it.

Focusing on the evaluation time of the hashing algorithms, the literature~\cite{DBLP:conf/sebd/ColuzziBL23} has emphasized, among the consistent hashing algorithms that achieve both load balance and strict monotonicity, AnchorHash~\cite{DBLP:journals/ton/MendelsonVBLKO21} as the fastest one, with DxHash~\cite{DBLP:journals/corr/abs-2107-07930} reaching almost equal performance. AnchorHash hashes in $O\left(\left( 1 + \log\left(a/n\right)\right)^2\right)$ time, where $a$ is the above-mentioned upper bound on the number of resources, that is, the initially set resource capacity. AnchorHash can be made artificially fast by choosing $a$ equal to $n$, but it is unlikely to meet such a configuration in practice. The authors of AnchorHash therefore benchmark their algorithm with $a/n = 1$, $1.1$, $2$ and $10$. As reported in Table~\ref{table:hashtimeconstantn}, we have found in our benchmarks that AnchorHash is about twice as fast as FlipHash with $a = n$, and about 3 times slower with $a = 10n$. Both are comparably fast for a value of $a \approx 2n$. This has been measured with 64-bit integer keys as input, and using as an underlying hash family for AnchorHash a bit-mixing construct that is similar to the one used for FlipHash.

\begin{table}[h]
  \centering
  \begin{tabular}{|c c c c|}
    \hline
    Algorithm  & $n$   & $a$      & Evaluation time \\ [0.5ex]
    \hline
    FlipHash   & $100$ & $2^{64}$ & $\qty{5.6}{ns}$ \\
    JumpHash   & $100$ & $2^{64}$ & $\qty{16}{ns}$  \\
    AnchorHash & $100$ & $100$    & $\qty{2.9}{ns}$ \\
    AnchorHash & $100$ & $110$    & $\qty{3.2}{ns}$ \\
    AnchorHash & $100$ & $200$    & $\qty{5.6}{ns}$ \\
    AnchorHash & $100$ & $1000$   & $\qty{13}{ns}$  \\
    \hline
  \end{tabular}
  \caption{Evaluation wall times of FlipHash, JumpHash, and AnchorHash with $n=100$ resources and various resource capacities $a$.}
  \label{table:hashtimeconstantn}
\end{table}

One challenge of working with AnchorHash is the need for setting a resource capacity $a$ that, while being high enough to accommodate for the maximum number of resources over the whole lifetime of the system, is chosen low enough so as to avoid an excessively high evaluation time due to a high ratio $a/n$. This is particularly problematic as modern distributed systems often process workloads that are highly variable over time, with future growth initially hard to predict. Systems scaled with workloads that have seasonal patterns will also inevitably cause AnchorHash to become less efficient when the workload is low. Table~\ref{table:hashtimeconstanta} shows that the constant evaluation time of FlipHash is a strong advantage, as it makes FlipHash's performance not only fast, but also predictable, stable, and independent of any initial configuration or the state of the system.

\begin{table}[h]
  \centering
  \begin{tabular}{|c c c c|}
    \hline
    Algorithm  & $n$    & $a$      & Evaluation time \\ [0.5ex]
    \hline
    FlipHash   & $10$   & $2^{64}$ & $\qty{6.1}{ns}$ \\
    FlipHash   & $100$  & $2^{64}$ & $\qty{5.7}{ns}$ \\
    FlipHash   & $1000$ & $2^{64}$ & $\qty{4.6}{ns}$ \\
    JumpHash   & $10$   & $2^{64}$ & $\qty{8.4}{ns}$ \\
    JumpHash   & $100$  & $2^{64}$ & $\qty{16}{ns}$  \\
    JumpHash   & $1000$ & $2^{64}$ & $\qty{25}{ns}$  \\
    AnchorHash & $10$   & $1000$   & $\qty{26}{ns}$  \\
    AnchorHash & $100$  & $1000$   & $\qty{13}{ns}$  \\
    AnchorHash & $1000$ & $1000$   & $\qty{2.9}{ns}$ \\
    \hline
  \end{tabular}
  \caption{Evaluation wall times of FlipHash, JumpHash, and AnchorHash with various numbers of resources $n$ and a resource capacity $a$ of at least $1000$.}
  \label{table:hashtimeconstanta}
\end{table}

\section{Conclusion}

With FlipHash, we introduce a new consistent range-hashing algorithm that achieves both monotonicity and regularity in constant time. We present FlipHash generically by building upon any family of hash functions with appropriate $(\epsilon, k)$-wise independence, and we also give a concrete implementation using either bit-mixing constructs or the XXH3 hash algorithm. We benchmark this implementation to demonstrate that not only does it empirically realize the properties of monotonicity and regularity that we formally prove, but that it also exceeds the computational performance of existing consistent range-hashing algorithms in all practical settings.

While this new algorithm can be used in a variety of applications that tolerate the constraint of sequentially indexing hash values, such as horizontal partitioning schemes, future work might focus on relaxing that constraint and applying ideas developed here to the more general consistent hashing problem in order to build an algorithm that perfectly balances the keyspace across hash values, while supporting arbitrary hash value removals. One such method would be to implement MementoHash~\cite{coluzzi2024mementohash} using FlipHash as the underlying consistent range-hashing algorithm instead of JumpHash. It would be interesting to see to what extent doing so leads to significant improvements in the evaluation performance of MementoHash.

\bibliographystyle{ACM-Reference-Format}
\bibliography{fliphash}


\begin{thebibliography}{20}


\ifx \showCODEN    \undefined \def \showCODEN     #1{\unskip}     \fi
\ifx \showDOI      \undefined \def \showDOI       #1{#1}\fi
\ifx \showISBNx    \undefined \def \showISBNx     #1{\unskip}     \fi
\ifx \showISBNxiii \undefined \def \showISBNxiii  #1{\unskip}     \fi
\ifx \showISSN     \undefined \def \showISSN      #1{\unskip}     \fi
\ifx \showLCCN     \undefined \def \showLCCN      #1{\unskip}     \fi
\ifx \shownote     \undefined \def \shownote      #1{#1}          \fi
\ifx \showarticletitle \undefined \def \showarticletitle #1{#1}   \fi
\ifx \showURL      \undefined \def \showURL       {\relax}        \fi
\providecommand\bibfield[2]{#2}
\providecommand\bibinfo[2]{#2}
\providecommand\natexlab[1]{#1}
\providecommand\showeprint[2][]{arXiv:#2}

\bibitem[Appleton and O'Reilly(2015)]%
        {DBLP:journals/corr/AppletonO15}
\bibfield{author}{\bibinfo{person}{Ben Appleton} {and} \bibinfo{person}{Michael
  O'Reilly}.} \bibinfo{year}{2015}\natexlab{}.
\newblock \showarticletitle{Multi-probe consistent hashing}.
\newblock \bibinfo{journal}{\emph{CoRR}}  \bibinfo{volume}{abs/1505.00062}
  (\bibinfo{year}{2015}).
\newblock
\showeprint[arXiv]{1505.00062}
\urldef\tempurl%
\url{http://arxiv.org/abs/1505.00062}
\showURL{%
\tempurl}


\bibitem[Collet(2019)]%
        {XXH3}
\bibfield{author}{\bibinfo{person}{Yann Collet}.}
  \bibinfo{year}{2019}\natexlab{}.
\newblock \bibinfo{title}{Presenting XXH3}.
\newblock
  \bibinfo{howpublished}{\url{https://fastcompression.blogspot.com/2019/03/presenting-xxh3.html}}.
\newblock


\bibitem[Coluzzi et~al\mbox{.}(2024)]%
        {coluzzi2024mementohash}
\bibfield{author}{\bibinfo{person}{Massimo Coluzzi}, \bibinfo{person}{Amos
  Brocco}, \bibinfo{person}{Alessandro Antonucci}, {and}
  \bibinfo{person}{Tiziano Leidi}.} \bibinfo{year}{2024}\natexlab{}.
\newblock \bibinfo{title}{MementoHash: A Stateful, Minimal Memory, Best
  Performing Consistent Hash Algorithm}.
\newblock
\newblock
\showeprint[arxiv]{2306.09783}~[cs.DC]


\bibitem[Coluzzi et~al\mbox{.}(2023)]%
        {DBLP:conf/sebd/ColuzziBL23}
\bibfield{author}{\bibinfo{person}{Massimo Coluzzi}, \bibinfo{person}{Amos
  Brocco}, {and} \bibinfo{person}{Tiziano Leidi}.}
  \bibinfo{year}{2023}\natexlab{}.
\newblock \showarticletitle{Consistently Faster: {A} Survey and Fair Comparison
  of Consistent Hashing Algorithms}. In \bibinfo{booktitle}{\emph{Proceedings
  of the 31st Symposium of Advanced Database Systems, Galzingano Terme, Italy,
  July 2nd to 5th, 2023}} \emph{(\bibinfo{series}{{CEUR} Workshop Proceedings},
  Vol.~\bibinfo{volume}{3478})}, \bibfield{editor}{\bibinfo{person}{Diego
  Calvanese}, \bibinfo{person}{Claudia Diamantini}, \bibinfo{person}{Guglielmo
  Faggioli}, \bibinfo{person}{Nicola Ferro}, \bibinfo{person}{Stefano
  Marchesin}, \bibinfo{person}{Gianmaria Silvello}, {and}
  \bibinfo{person}{Letizia Tanca}} (Eds.). \bibinfo{publisher}{CEUR-WS.org},
  \bibinfo{pages}{51--64}.
\newblock
\urldef\tempurl%
\url{https://ceur-ws.org/Vol-3478/paper03.pdf}
\showURL{%
\tempurl}


\bibitem[DeCandia et~al\mbox{.}(2007)]%
        {DeCandia2007}
\bibfield{author}{\bibinfo{person}{Giuseppe DeCandia}, \bibinfo{person}{Deniz
  Hastorun}, \bibinfo{person}{Madan Jampani}, \bibinfo{person}{Gunavardhan
  Kakulapati}, \bibinfo{person}{Avinash Lakshman}, \bibinfo{person}{Alex
  Pilchin}, \bibinfo{person}{Swaminathan Sivasubramanian},
  \bibinfo{person}{Peter Vosshall}, {and} \bibinfo{person}{Werner Vogels}.}
  \bibinfo{year}{2007}\natexlab{}.
\newblock \showarticletitle{Dynamo: Amazon’s highly available key-value
  store}. In \bibinfo{booktitle}{\emph{ACM Symposium on Operating System
  Principles}}.
\newblock
\urldef\tempurl%
\url{https://www.amazon.science/publications/dynamo-amazons-highly-available-key-value-store}
\showURL{%
\tempurl}


\bibitem[Dong and Wang(2021)]%
        {DBLP:journals/corr/abs-2107-07930}
\bibfield{author}{\bibinfo{person}{Chaos Dong} {and} \bibinfo{person}{Fang
  Wang}.} \bibinfo{year}{2021}\natexlab{}.
\newblock \showarticletitle{DxHash: {A} Scalable Consistent Hash Based on the
  Pseudo-Random Sequence}.
\newblock \bibinfo{journal}{\emph{CoRR}}  \bibinfo{volume}{abs/2107.07930}
  (\bibinfo{year}{2021}).
\newblock
\showeprint[arXiv]{2107.07930}
\urldef\tempurl%
\url{https://arxiv.org/abs/2107.07930}
\showURL{%
\tempurl}


\bibitem[Eisenbud et~al\mbox{.}(2016)]%
        {DBLP:conf/nsdi/EisenbudYCSKMCC16}
\bibfield{author}{\bibinfo{person}{Daniel~E. Eisenbud}, \bibinfo{person}{Cheng
  Yi}, \bibinfo{person}{Carlo Contavalli}, \bibinfo{person}{Cody Smith},
  \bibinfo{person}{Roman Kononov}, \bibinfo{person}{Eric Mann{-}Hielscher},
  \bibinfo{person}{Ardas Cilingiroglu}, \bibinfo{person}{Bin Cheyney},
  \bibinfo{person}{Wentao Shang}, {and} \bibinfo{person}{Jinnah~Dylan Hosein}.}
  \bibinfo{year}{2016}\natexlab{}.
\newblock \showarticletitle{Maglev: {A} Fast and Reliable Software Network Load
  Balancer}. In \bibinfo{booktitle}{\emph{13th {USENIX} Symposium on Networked
  Systems Design and Implementation, {NSDI} 2016, Santa Clara, CA, USA, March
  16-18, 2016}}, \bibfield{editor}{\bibinfo{person}{Katerina~J. Argyraki} {and}
  \bibinfo{person}{Rebecca Isaacs}} (Eds.). \bibinfo{publisher}{{USENIX}
  Association}, \bibinfo{pages}{523--535}.
\newblock
\urldef\tempurl%
\url{https://www.usenix.org/conference/nsdi16/technical-sessions/presentation/eisenbud}
\showURL{%
\tempurl}


\bibitem[Evensen(2019)]%
        {Moremur}
\bibfield{author}{\bibinfo{person}{Pelle Evensen}.}
  \bibinfo{year}{2019}\natexlab{}.
\newblock \bibinfo{title}{Stronger, better, morer, Moremur; a better
  Murmur3-type mixer}.
\newblock
  \bibinfo{howpublished}{\url{https://mostlymangling.blogspot.com/2019/12/stronger-better-morer-moremur-better.html}}.
\newblock


\bibitem[Karger et~al\mbox{.}(1997)]%
        {DBLP:conf/stoc/KargerLLPLL97}
\bibfield{author}{\bibinfo{person}{David~R. Karger}, \bibinfo{person}{Eric
  Lehman}, \bibinfo{person}{Frank~Thomson Leighton}, \bibinfo{person}{Rina
  Panigrahy}, \bibinfo{person}{Matthew~S. Levine}, {and}
  \bibinfo{person}{Daniel Lewin}.} \bibinfo{year}{1997}\natexlab{}.
\newblock \showarticletitle{Consistent Hashing and Random Trees: Distributed
  Caching Protocols for Relieving Hot Spots on the World Wide Web}. In
  \bibinfo{booktitle}{\emph{Proceedings of the Twenty-Ninth Annual {ACM}
  Symposium on the Theory of Computing, El Paso, Texas, USA, May 4-6, 1997}},
  \bibfield{editor}{\bibinfo{person}{Frank~Thomson Leighton} {and}
  \bibinfo{person}{Peter~W. Shor}} (Eds.). \bibinfo{publisher}{{ACM}},
  \bibinfo{pages}{654--663}.
\newblock
\urldef\tempurl%
\url{https://doi.org/10.1145/258533.258660}
\showDOI{\tempurl}


\bibitem[Karger et~al\mbox{.}(1999)]%
        {DBLP:journals/cn/KargerSBBDIKMY99}
\bibfield{author}{\bibinfo{person}{David~R. Karger}, \bibinfo{person}{Alex
  Sherman}, \bibinfo{person}{Andy Berkheimer}, \bibinfo{person}{Bill Bogstad},
  \bibinfo{person}{Rizwan Dhanidina}, \bibinfo{person}{Ken Iwamoto},
  \bibinfo{person}{Brian Kim}, \bibinfo{person}{Luke Matkins}, {and}
  \bibinfo{person}{Yoav Yerushalmi}.} \bibinfo{year}{1999}\natexlab{}.
\newblock \showarticletitle{Web Caching with Consistent Hashing}.
\newblock \bibinfo{journal}{\emph{Comput. Networks}} \bibinfo{volume}{31},
  \bibinfo{number}{11-16} (\bibinfo{year}{1999}), \bibinfo{pages}{1203--1213}.
\newblock
\urldef\tempurl%
\url{https://doi.org/10.1016/S1389-1286(99)00055-9}
\showDOI{\tempurl}


\bibitem[Lakshman and Malik(2010)]%
        {DBLP:journals/sigops/LakshmanM10}
\bibfield{author}{\bibinfo{person}{Avinash Lakshman} {and}
  \bibinfo{person}{Prashant Malik}.} \bibinfo{year}{2010}\natexlab{}.
\newblock \showarticletitle{Cassandra: a decentralized structured storage
  system}.
\newblock \bibinfo{journal}{\emph{{ACM} {SIGOPS} Oper. Syst. Rev.}}
  \bibinfo{volume}{44}, \bibinfo{number}{2} (\bibinfo{year}{2010}),
  \bibinfo{pages}{35--40}.
\newblock
\urldef\tempurl%
\url{https://doi.org/10.1145/1773912.1773922}
\showDOI{\tempurl}


\bibitem[Lamping and Veach(2014)]%
        {DBLP:journals/corr/LampingV14}
\bibfield{author}{\bibinfo{person}{John Lamping} {and} \bibinfo{person}{Eric
  Veach}.} \bibinfo{year}{2014}\natexlab{}.
\newblock \showarticletitle{A Fast, Minimal Memory, Consistent Hash Algorithm}.
\newblock \bibinfo{journal}{\emph{CoRR}}  \bibinfo{volume}{abs/1406.2294}
  (\bibinfo{year}{2014}).
\newblock
\showeprint[arXiv]{1406.2294}
\urldef\tempurl%
\url{http://arxiv.org/abs/1406.2294}
\showURL{%
\tempurl}


\bibitem[Maiga(2020)]%
        {BitMixer}
\bibfield{author}{\bibinfo{person}{Jon Maiga}.}
  \bibinfo{year}{2020}\natexlab{}.
\newblock \bibinfo{title}{The construct of a bit mixer}.
\newblock
  \bibinfo{howpublished}{\url{https://jonkagstrom.com/bit-mixer-construction/}}.
\newblock


\bibitem[Mendelson et~al\mbox{.}(2021)]%
        {DBLP:journals/ton/MendelsonVBLKO21}
\bibfield{author}{\bibinfo{person}{Gal Mendelson}, \bibinfo{person}{Shay
  Vargaftik}, \bibinfo{person}{Katherine Barabash}, \bibinfo{person}{Dean~H.
  Lorenz}, \bibinfo{person}{Isaac Keslassy}, {and} \bibinfo{person}{Ariel
  Orda}.} \bibinfo{year}{2021}\natexlab{}.
\newblock \showarticletitle{AnchorHash: {A} Scalable Consistent Hash}.
\newblock \bibinfo{journal}{\emph{{IEEE/ACM} Trans. Netw.}}
  \bibinfo{volume}{29}, \bibinfo{number}{2} (\bibinfo{year}{2021}),
  \bibinfo{pages}{517--528}.
\newblock
\urldef\tempurl%
\url{https://doi.org/10.1109/TNET.2020.3039547}
\showDOI{\tempurl}


\bibitem[Mirrokni et~al\mbox{.}(2018)]%
        {DBLP:conf/soda/MirrokniTZ18}
\bibfield{author}{\bibinfo{person}{Vahab~S. Mirrokni}, \bibinfo{person}{Mikkel
  Thorup}, {and} \bibinfo{person}{Morteza Zadimoghaddam}.}
  \bibinfo{year}{2018}\natexlab{}.
\newblock \showarticletitle{Consistent Hashing with Bounded Loads}. In
  \bibinfo{booktitle}{\emph{Proceedings of the Twenty-Ninth Annual {ACM-SIAM}
  Symposium on Discrete Algorithms, {SODA} 2018, New Orleans, LA, USA, January
  7-10, 2018}}, \bibfield{editor}{\bibinfo{person}{Artur Czumaj}} (Ed.).
  \bibinfo{publisher}{{SIAM}}, \bibinfo{pages}{587--604}.
\newblock
\urldef\tempurl%
\url{https://doi.org/10.1137/1.9781611975031.39}
\showDOI{\tempurl}


\bibitem[Nakatani(2021)]%
        {DBLP:journals/tpds/Nakatani21}
\bibfield{author}{\bibinfo{person}{Yuichi Nakatani}.}
  \bibinfo{year}{2021}\natexlab{}.
\newblock \showarticletitle{Structured Allocation-Based Consistent Hashing With
  Improved Balancing for Cloud Infrastructure}.
\newblock \bibinfo{journal}{\emph{{IEEE} Trans. Parallel Distributed Syst.}}
  \bibinfo{volume}{32}, \bibinfo{number}{9} (\bibinfo{year}{2021}),
  \bibinfo{pages}{2248--2261}.
\newblock
\urldef\tempurl%
\url{https://doi.org/10.1109/TPDS.2021.3058963}
\showDOI{\tempurl}


\bibitem[Olteanu et~al\mbox{.}(2018)]%
        {DBLP:conf/nsdi/OlteanuAVR18}
\bibfield{author}{\bibinfo{person}{Vladimir~Andrei Olteanu},
  \bibinfo{person}{Alexandru Agache}, \bibinfo{person}{Andrei Voinescu}, {and}
  \bibinfo{person}{Costin Raiciu}.} \bibinfo{year}{2018}\natexlab{}.
\newblock \showarticletitle{Stateless Datacenter Load-balancing with Beamer}.
  In \bibinfo{booktitle}{\emph{15th {USENIX} Symposium on Networked Systems
  Design and Implementation, {NSDI} 2018, Renton, WA, USA, April 9-11, 2018}},
  \bibfield{editor}{\bibinfo{person}{Sujata Banerjee} {and}
  \bibinfo{person}{Srinivasan Seshan}} (Eds.). \bibinfo{publisher}{{USENIX}
  Association}, \bibinfo{pages}{125--139}.
\newblock
\urldef\tempurl%
\url{https://www.usenix.org/conference/nsdi18/presentation/olteanu}
\showURL{%
\tempurl}


\bibitem[Sackman(2015)]%
        {DBLP:journals/corr/Sackman15}
\bibfield{author}{\bibinfo{person}{Matthew Sackman}.}
  \bibinfo{year}{2015}\natexlab{}.
\newblock \showarticletitle{Perfect Consistent Hashing}.
\newblock \bibinfo{journal}{\emph{CoRR}}  \bibinfo{volume}{abs/1503.04988}
  (\bibinfo{year}{2015}).
\newblock
\showeprint[arXiv]{1503.04988}
\urldef\tempurl%
\url{http://arxiv.org/abs/1503.04988}
\showURL{%
\tempurl}


\bibitem[Thaler and Ravishankar(1996)]%
        {Rendezvous}
\bibfield{author}{\bibinfo{person}{David Thaler} {and}
  \bibinfo{person}{Chinya~V. Ravishankar}.} \bibinfo{year}{1996}\natexlab{}.
\newblock \showarticletitle{A Name-Based Mapping Scheme for Rendezvous}.
\newblock \bibinfo{journal}{\emph{Technical Report CSE-TR-316-96, University of
  Mishigan}} (\bibinfo{year}{1996}).
\newblock
\urldef\tempurl%
\url{https://www.eecs.umich.edu/techreports/cse/96/CSE-TR-316-96.pdf}
\showURL{%
\tempurl}


\bibitem[Thaler and Ravishankar(1998)]%
        {DBLP:journals/ton/ThalerR98}
\bibfield{author}{\bibinfo{person}{David Thaler} {and}
  \bibinfo{person}{Chinya~V. Ravishankar}.} \bibinfo{year}{1998}\natexlab{}.
\newblock \showarticletitle{Using name-based mappings to increase hit rates}.
\newblock \bibinfo{journal}{\emph{{IEEE/ACM} Trans. Netw.}}
  \bibinfo{volume}{6}, \bibinfo{number}{1} (\bibinfo{year}{1998}),
  \bibinfo{pages}{1--14}.
\newblock
\urldef\tempurl%
\url{https://doi.org/10.1109/90.663936}
\showDOI{\tempurl}


\end{thebibliography}

\end{document}